\documentclass[%
 reprint,
superscriptaddress,
 amsmath,amssymb,
 aps,
]{revtex4-2}

\usepackage{float}
\usepackage[svgnames]{xcolor}

\usepackage{orcidlink}
\usepackage{euscript,graphicx}

\usepackage{autobreak,cleveref}
\usepackage{amsthm,dsfont,amsfonts,amsmath,amssymb,wasysym}
\usepackage{euscript,color,fontenc,textcomp,relsize}
\usepackage{bm,url,float}

\makeatletter
\DeclareUnicodeCharacter{2212}{\textminus}
\usepackage{graphicx}
\usepackage{dcolumn}
\usepackage{bm}

\hypersetup{
    colorlinks=true,
    citecolor=green,
    linkcolor=black,
    urlcolor=blue
}

\begin{document}


\title{On Constraining the Proposed Hierarchical Triple Scenario for an Eccentric Milli-Second Pulsar Binary PSR J1618--3921}%

\newcommand{\IITR}{Department of Physics, Indian Institute of Technology Roorkee\\
Roorkee, Uttarakhand, India.}

\author{Adya Shukla\orcidlink{0009-0005-7058-5539}}
 \email{adyashukla0496@gmail.com}
\affiliation{\IITR
}%

\author{A.~Gopakumar\orcidlink{0000-0003-4274-4369}}
 \email{gopu.tifr@gmail.com}
\affiliation{%
 Department of Astronomy and Astrophysics, Tata Institute of Fundamental Research\\
 Mumbai, Maharashtra, India.
}%

\author{Paramasivan Arumugam\orcidlink{https://orcid.org/0000-0001-9624-8024}}
\affiliation{\IITR
}%


\begin{abstract}
A very recent and meticulous timing effort suggests that an eccentric millisecond pulsar  (eMSP) binary, namely PSR J1618--3921, is likely to be a part of a hierarchical triple (HT) system 
with a $0.6M_{\odot}$ companion in a $\sim 300$yr orbit. We investigate observational implications of the proposed HT scenario for PSR J1618--3921 and our ability to constrain the scenario. We model the MSP-Helium White Dwarf binary to be a part of bound point-mass HT, while incorporating the effects due to the quadrupolar interactions between the inner and outer binaries, along with dominant order general relativistic contributions to the periastron precession of the inner binary.
If the proposed HT system is indeed undergoing Kozai oscillations at the present epoch, the orbital eccentricity ($e$) would be expected to decrease, while the rate of periastron advance ($\dot{\omega}$) would correspondingly increase, for plausible ranges of the HT parameters. Furthermore, the fractional variations in $e$ are anticipated to be at the level of a few parts in $10^{5}$—a magnitude that is substantially larger than the current measurement precision of $e$.
We find that, for this eccentric MSP binary, the HT configurations that minimize the temporal evolution of orbital eccentricity and argument of periastron are mutually incompatible. 
This indicates that the continued high-precision timing of PSR J1618–3921—when analyzed within the framework introduced here—should place stringent limits on the presence and properties of a potential third body in the system.

\end{abstract}

\maketitle

\section{Introduction}\label{sec1:level1}

Millisecond pulsar (MSP) binaries typically host rapidly spinning neutron stars and exhibit extremely small orbital eccentricities \cite{Alpar}. Their formation scenario involves a phase where a slowly rotating neutron star accretes material from its companion star and the inherent strong tidal interactions circularize the orbit as material flows steadily onto the neutron star, and spinning it up to millisecond periods \cite{taurisheuvel}. It is expected that such a process lasts millions of years, efficiently erasing any initial orbital eccentricity. Once the mass transfer ceases and the companion becomes a white dwarf (WD), the orbit remains locked in this highly circular configuration as no significant mass loss or dynamical interactions are expected to occur in such a nascent MSP-WD binary~\cite{Phinney_1992}. The fraction of MSP binaries in nearly circular orbits is $ >90\% $, and it reflects the dominant role of long-term stable mass transfer and tidal forces in shaping these systems \cite{Lorimer_2008}. 
\par 
Accreting Millisecond X-ray Pulsars (AMXPs) provide a direct observational link between ordinary neutron stars and recycled MSPs~\cite{Patruno_2012}. Their spin periods, accretion behavior, and circular orbits provide strong, real-time evidence supporting the idea that MSPs form in binaries through accretion, and that this process naturally circularizes their orbits \cite{Patruno_2020}.
Further, a very recent effort provides strong observational evidence for the occurrence of the common envelope phase in PSR J1928+1815, which is essentially critical in the 
formation of tight MSP binaries — especially those with WD companions in short-period, and nearly circular orbits \cite{Yang_2025}. 
\par 
These considerations ensure that rare
eccentric millisecond pulsar (eMSP) binaries are fascinating systems that deviate from the typical low-eccentricity MSP population \cite{Freire2011}.
Unlike the standard formation pathway, where long-term mass transfer from a companion star circularizes the orbit, these systems retain or acquire significant eccentricity (e $>$ 0.01). Most known eccentric MSP binaries are found in dense stellar environments like globular clusters, where dynamical interactions, such as exchange encounters or close flybys, can perturb otherwise circular orbits \cite{Ransom_2005,Wilson_2008}. However, there exists a smaller number of eMSP binaries in the Galactic field with orbital eccentricities ranging from 0.03 to 0.1, and their orbital periods are typically in days.
Notable examples include J1903+0327 \cite{Champion_2008,Freire2011}, J1618--3921 \cite{Edwards_2001,Octau_2018}, J1950+2414 \cite{Knispel_2015,Zhu_2019}, J2234+0611 \cite{Deneva_2013,Stovall_2019}, J1946+3417 \cite{Barr_2013,Barr_2017}, J1146--6610
\cite{Lorimer_2021} and J0955--6150 \cite{Serylak_2022}.
These eMSPs are very intriguing because they typically possess helium white dwarf (HeWD) companions and adhere to the standard mass–orbital period relationship observed in typical MSP binaries \cite{tauris1999}.
In contrast, 
PSR J1903+0327 is a massive millisecond pulsar in a highly eccentric $(e \sim 0.4)$
and wide 
orbit ($P_b\sim95$ days) with a 
main-sequence companion \cite{Freire2011,Khargharia_2012}.

\par 
The presence of WD companions in eMSP binaries suggests that they are likely formed through conventional Roche lobe overflow (RLO) mass transfer processes.
Therefore, the field eMSP binaries might have acquired their eccentricities through intrinsic evolutionary processes~\cite{Antoniadis_2014},  
triple-star evolution~\cite{Kozai_1962}, or the accretion-induced collapse of a white dwarf~\cite{Friere_Tauris_2014}. 
In this paper, we take a detailed look at a carefully crafted proposal that 
PSR J1618--3921 binary system is likely part of a hierarchical triple (HT)\cite{Grunthal_2024}.
Recall that a HT involves three objects where two of them form a close (inner) binary, and the third object orbits the center of mass of that binary at a larger distance \cite{Toonen16}.

\par 
The HT scenario for PSR J1618--3921 is based on
a comprehensive and detailed timing analysis of the pulsar, pursued by \cite{Grunthal_2024},  using 23 years of data from MeerKAT, Parkes, and Nançay radio telescopes. A set of thorough investigations and the measurement of periastron advance 
 allowed \cite{Grunthal_2024} to infer accurately both masses of the eMSP binary.
The pulsar mass is found to be
$ m_p = 1.20^{+0.19}_{-0.20} M_\odot$ while the companion mass is $m_c = 0.20^{+0.11}_{-0.03}M_\odot$, which suggests that the companion is a HeWD.
The inferences on these two masses and the orbital inclination were possible as 
\cite{Grunthal_2024} analysed the final dataset using the timing software package \texttt{TEMPO2}~\cite{TEMPO2} and 
employed a   
Bayesian timing analysis framework, namely \texttt{TEMPONEST}~\cite{TEMPONEST} to fit for the Shapiro parameters and derived posterior distributions for eMSP binary's orbital inclination, $m_p$ and $m_c$.
More interestingly, \cite{Grunthal_2024} measured 
an anomalously large and negative orbital period derivative ($ \dot P_b \sim -2.26^{+0.35}_{-0.33}\times\, 10^{-12}$) and a significant second derivative of eMSP's spin frequency ($\ddot f \sim -1.0(2)~\times10^{-27} s^{-3}$), thanks to the high-quality MeerTime data.
It turns out that these observations could not be explained by known effects such as Galactic acceleration or the Shklovskii effect
\cite{Phinney_1992,Shklovskii_1970,MWac}. 
However, the observed anomalous timing acceleration could be explained by the presence of 
a third body having a mass of $\sim 0.6 M_{\odot}$ with an orbital period of $\sim 300$ years 
\cite{Grunthal_2024, JR_tb}.
Further, a careful exploration for a possible optical counterpart in the 
\texttt{DECaPS2} catalogue~\cite{Schlafly_2018} suggests that the third body could be
 a  M-dwarf with mass $ <0.56 M_{\odot}$ or a
compact object with an absolute magnitude $>9.79$ mag~\cite{Grunthal_2024}.
These considerations 
 compelled \cite{Grunthal_2024} to propose that PSR J1618--3921 is part of a hierarchical triple (HT) system where the inner binary consists of 
$\sim 12ms$ MSP and its HeWD companion in a $\sim 22$ day orbit with an eccentricity of $\sim 0.03$ while the third body orbits the center of mass of the (inner) binary with an orbital period of $\sim 300$ years.
\par

  This paper explores the possible observational implications of the proposed HT scenario for  PSR J1618--3921, influenced by \cite{Gopakumar_2009}.
 We begin by speculating that the proposed HT configuration may be experiencing 
 Kozai oscillations, driven
 by the fact that this compact MSP-HeWD binary exhibits 
 a large orbital eccentricity of $0.0274$, unusual for such binaries.
This mechanism, referred to as von Zeipel-Kozai-Lidov, Lidov–Kozai, Kozai–Lidov effect, oscillation, cycle, or resonance \cite{vonZeipel1910,Naoz_2016,Kozai_1962,Lidov_1962},  is 
a dynamical effect that occurs in HT systems, where a distant third body gravitationally perturbs a compact inner binary in an inclined orbit. This resonance can cause periodic oscillations in the inner binary's eccentricity and argument of periastron while conserving angular momentum by redistributing it between the two orbital planes \cite{Blaes_2002}. 
\par 
In this paper, 
we argue that the existing and ongoing timing campaign should be able to rule out the presence of Kozai resonance in the eMSP binary PSR J1618--3921.
However, it is possible that the proposed HT system is not currently experiencing 
Kozai resonance, which prompted us to explore the observational implications of 
PSR J1618--3921 is part of a HT, influenced by the inferences of Ref.~\cite{Grunthal_2024}.
We restrict our attention to those third-body  configurations that {\it minimize}
its effects on the temporal evolution of the inner binary's orbital eccentricity and argument of periastron. 

Our numerical experiments suggest that the ongoing timing of PSR J1618--3921 should allow 
one to rule out the proposed HT scenario for this eMSP binary, especially by 
constraining any evolution of the orbital eccentricity and the rate of 
periastron advance.

\par 
In what follows, we summarize a mathematical formalism useful for modeling such HT configurations with the help of \cite{Blaes_2002}.
In Sec.~\ref{sec3:level1}, we explore the observational implications of 
such a scenario on the temporal evolution of measurable timing parameters for PSR J1618--3921.
Our conclusions and discussions are presented in Sec.~\ref{sec4}.

\section{\label{sec2:level1}Modeling the Proposed Hierarchical Triple Configuration for PSR J618--3921}
 We treat the proposed HT in PSR~J1618--3921 as consisting of two interacting eccentric 
 binaries whose Keplerian orbits are inclined to each other.
The inner binary consists of PSR J1618--3921 and its HeWD companion, and we let their 
masses to be $m_0$ and $m_1$.  
The outer binary consists of an M-dwarf/compact object of mass $m_2$  with 
the inner binary’s center of mass. Motivated by~\cite{Blaes_2002}, 
we denote the orbital elements of inner and outer binaries, namely their eccentricities and semi-major axes, to be ($e_1$, $e_2$), and 
($a_1$, $a_2$), respectively. Additionally, we let ($g_1$, $g_2$) be 
their respective arguments of periastron, which are measured with respect to their own line of nodes,
and the angle $\iota$ provides 
 the mutual angle of inclination between these two orbital planes.
 We now have the relevant variables to investigate the temporal evolution of the HT system 
 involving PSR~J1618--3921.
\par

 We employ 
 the secular perturbation theory to describe Newtonian HT systems, detailed in \cite{Blaes_2002,Ford_review},
 to evolve the inner binary's eccentricity and argument of periastron. 
This approach  provides differential equations that describe 
the orbital averaged secular evolution of $(e_1,g_1)$ while ignoring orbital time scale variations in these variables. 
We incorporate only the dominant order gravitational interactions between the inner and outer orbits, and it 
provides corrections to Newtonian two-body dynamics in terms of  
the small parameter $\alpha = (a_1/a_2)$. The expressions describing the secular temporal evolution of the inner binary’s eccentricity and argument of periastron as detailed in~\cite{Blaes_2002,Gopakumar_2009} are given by

\begin{subequations} 
\label{etgt_Eq}
\begin{align}
\label{dgdt}
\frac{d g_1}{dt} &= 
\frac{ 6\,{\cal C}_2}{{\cal G}_1} \biggl \{ 4\theta^2+(5\cos2g_1-1)(1-e_1^2-
                 \theta^2) \biggr \} 
\nonumber\\
 &
+\frac { 6\, {\cal C}_2\, \theta }{  {\cal G}_2} \biggl \{ 2+e_1^2(3-5\cos2g_1) \biggr \} 
\nonumber\\
 &
+{3\over c^2\, a_1\, (1-e_1^2)}\,
\biggl [ \frac {G\, M_i }{ a_1} \biggr ]^{3/2}
\,,\\
\label{dedt}
\frac{d e_1}{dt} & =\frac{ 30\, {\cal C}_2}{ {\cal G}_1} \, {e_1(1-e_1^2)}\, (1-\theta^2)\, \sin 2g_1
\,,
\end{align}
\end{subequations}
where $M_i = (m_0 + m_1)$ is the total mass of the inner binary and $\theta = \cos \iota $. The following expressions define ${\cal C}_2$ and the angular momenta ${\cal G}_1$ and ${\cal G}_2$ associated with the inner and outer binaries:

\begin{subequations} 
\label{c2g12_Eq}
\begin{align}
\label{c2_def}
{\cal C}_2 &=  \frac{ G\, M_i\, \eta_i} {16\, a_2} \, \frac{ m_2}{ (1-e_2^2)^{3/2} } \, 
{\alpha}^2
\,,\\
{\cal G}_1 &= \eta_i\, \biggl \{ G\, M_i^3\, a_1\, (1-e_1^2) \biggr \}^{1/2} 
\,,\\
{\cal G}_2 &=  m_2 \biggl \{ 
\frac{ G\, M_i^2}{ ( M_i + m_2) } \, a_2\, (1-e_2^2) 
\biggr \}^{1/2} 
\,,
\end{align}
\end{subequations}
where $\eta_i$, the symmetric mass ratio of the inner binary, is given by $\eta_i = m_0m_1/M_i^2$. In these equations, we supplement the classical quadrupolar contributions to $\dot g_1$ with the general relativistic evolution for $g_1$ in an ad hoc manner, influenced by \cite{Blaes_2002}.  In particular, the third term in Eq.~\ref{dgdt} represents the leading-order post-Newtonian (1PN) contribution to $\dot{g}_1$, which provides general relativistic corrections to the Newtonian binary dynamics. Note that these corrections are of the order of  $(v/c)^2 \sim G M_i / (c^2 r)$, where $r$, $v$, and $M_i$ denote the orbital separation, relative velocity, and total mass of the inner binary \cite{BS89}.

\par 
 A few comments are in order. It should be obvious that we have not
 incorporated the dominant order general relativistic contributions to  $ \dot e_1$ which 
arises from the gravitational wave emission induced losses to the inner binary's 
orbital energy and angular momentum \cite{BS89}.
This is justified as such general relativistic contributions to $d e_1/dt$ appear at the $(v/c)^5$ order compared to  $(v/c)^2$ order for $d g_1/dt$. 
Note that the above-listed quadrupolar order Newtonian contributions 
to $(d g_1/dt,d e_1/dt)$ is obtainable from certain 
`doubly averaged' Hamiltonian, which is derivable from the usual Hamiltonian for an HT at the quadrupolar interaction order \cite{Ford_review}.
The `doubly averaged' Hamiltonian, suitable for describing secular (long-term) temporal evolution of an HT, is independent of the mean anomalies of the inner and outer orbits. This implies that their respective conjugate momenta and hence the semi-major axes $(a_1,a_2)$ are constants of motion. 
\par 
Interestingly, when one neglects 
general relativistic contributions and lets ${\cal G}_2 >> {\cal G}_1$ in the above equations, 
it leads to an analytic solution \cite{Kozai_1962}. Further, 
it is possible to construct an approximate integral of motion in terms of $e_1$ and $\theta$ that allows one to classify the dynamical behavior of a HT with the help of trajectories in the phase space defined by $e_1$ and  $\cos g_1$. It turns out that if the mutual inclination angle is fairly high and in a certain window, the orbital eccentricities experience periodic oscillations over time-scales that are extremely large compared to the respective orbital periods. The above effect, usually referred to as the 
Kozai resonance/von Zeipel-Kozai-Lidov oscillations~\cite{vonZeipel1910,Naoz_2016,Kozai_1962,Lidov_1962} arise due to the tidal torquing between the two orbits. The Kozai resonance can force initially tiny eccentricity of the inner binary to oscillate through a maximum value, given by ${e_1}^{\rm max} \simeq \left( 1 - \frac{5}{3}\, \cos^2 \iota_{0} \right)^{1/2} $ where $\iota_{0}$ is the initial value for the mutual inclination angle.
Due to an expected restriction, namely  $| \cos \iota_{0} <  (3/5)^{1/2} |$, $\iota_{0}$ is required  to lie in the range $39^{\circ} - 141^{\circ}$ \cite{Blaes_2002}.
This is essentially why we employ Eqs.~\ref{etgt_Eq} and~\ref{c2g12_Eq}
 to probe if PSR~J1618--3921 is 
part of a HT that experiences Kozai resonance, though these equations are valid for other HT configurations \cite{Blaes_2002}.
\par 
We now list ways to obtain 
possible bounds on $\alpha = (a_1/a_2)$, critical to implement Eqs.~\ref{etgt_Eq} and~\ref{c2g12_Eq},
while employing the timing results for PSR~J1618–3921, listed as 
Table. 3 in \cite{Grunthal_2024}. A lower $\alpha$ bound arises from the possibility that the general relativistic periastron advance of the inner binary could interfere with the Kozai resonance and even terminate the eccentricity oscillations~\cite{Blaes_2002}. 
The following bound on $\alpha$ ensures that this does not 
occur as argued in \cite{Blaes_2002}:
\begin{equation}
\label{alpha_GR}
 \left ( \frac{ a_2}{a_1} \right ) < \biggl [ \frac{3}{4}\, \frac{ m_2 }{M_i} \, \frac {\tilde a_1} { \tilde M_i } \,
\left ( \frac{ 1 - e_1^2 } { 1 -e_2^2}  \right )^{3/2}  \biggr ]^{1/3} 
\,,
\end{equation}
\newline
where $ \tilde M_i $ is $M_i/M_{\odot}, \tilde a_1 = a_1/L_{\odot} $ and $ L_{\odot} = 1.476625$ km. The above inequality is obtained by equating the right-hand side of Eq.~(\ref{dgdt}) to zero, after neglecting the much smaller  ${\cal C}_2/{\cal G}_2$ contributions to 
$d g_1/dt$ and demanding that the resulting expression for $ \cos^2 \iota$ remains positive. 
\par 
A possible upper bound on $\alpha$ arises by employing an empirical relation that is relevant while discussing the stability of Newtonian coplanar prograde orbits in HT configurations, detailed in \cite{MA_2001}, and it reads
\begin{equation}
\label{alpha_MA}
 \left ( \frac{ a_2}{a_1} \right ) >  \frac{ 2.8} { 1 -e_2} \, \biggl [  \biggl ( 1 + \frac{ m_2}{M_i} \biggr )\, 
\frac{ ( 1 + e_2) }{ ( 1-e_2)^{1/2} } \biggr ] ^{2/5}
\,.
\end{equation}
It is advisable to treat the above inequality to be rather conservative as the inclined orbits involved in HT configurations are expected to be more stable than the coplanar triples of Eq.~(\ref{alpha_MA}), as noted in \cite{MA_2001}.

\begin{figure*}[!ht]
    \centering
    \includegraphics[width=0.79\textwidth]{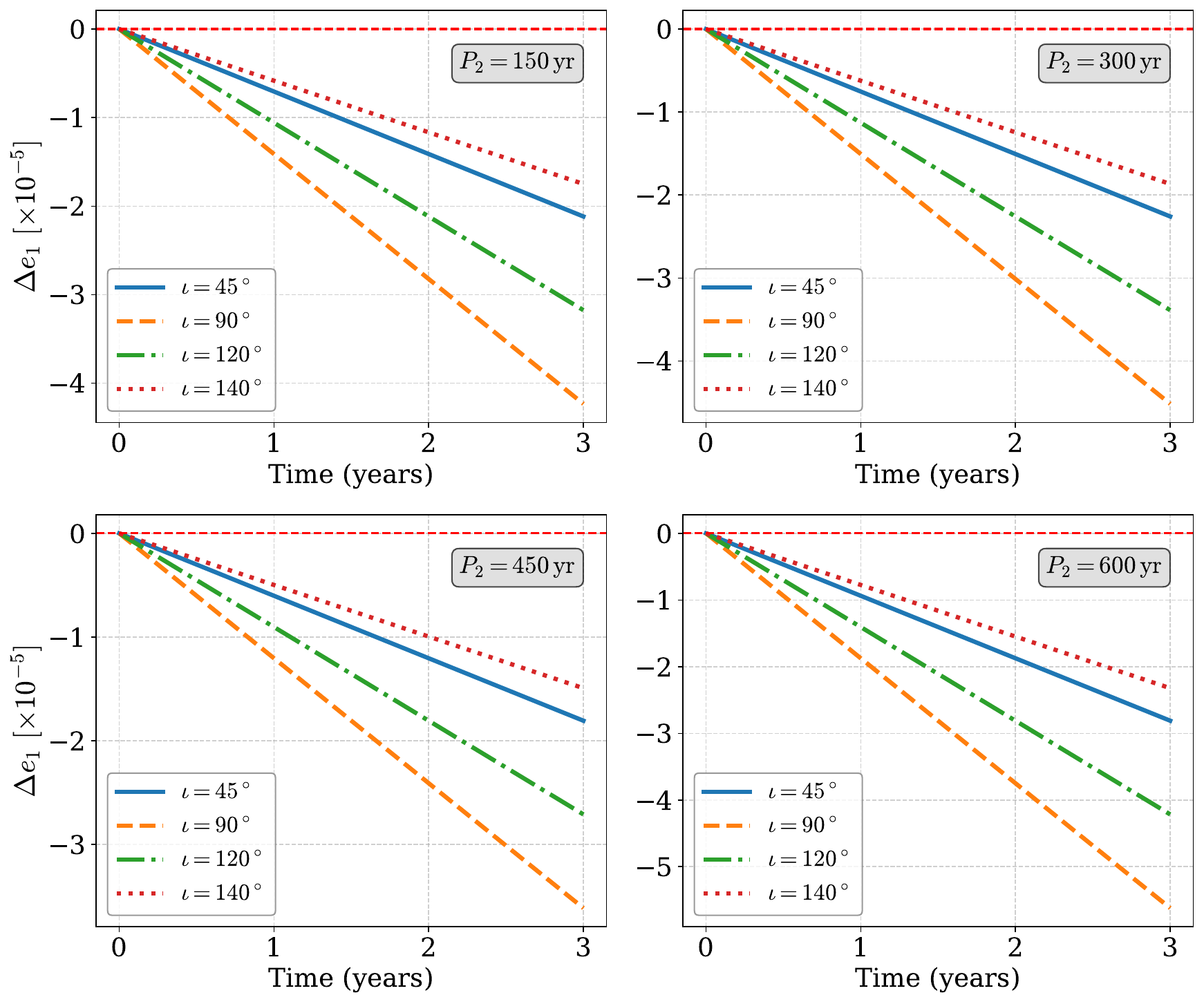}
    \caption{\label{fig:frac_change_e1}
Plots showing the temporal evolution of fractional changes in our eMSP binary's orbital eccentricity,
namely $ \Delta e = ( e_1(t) - e_1)/e_1$, if PSR J1618--3921 is part of a HT that experiences the Kozai resonance at the present epoch. The four panels are for HT configurations when the third body perturbations are characterized by $ \alpha = (a_1/a_2) = $ 1/200, 1/320, 1/420, and 1/510, and we vary the inclination
$\iota$ between the inner and outer orbits in the allowed region: [$39^\circ$ and $141^\circ$].
It should be obvious that $\Delta e(t) \sim 10^{-5}$ and such variations are substantially larger 
than  the observational precision of $10^{-8}$ for $e$ as reported in \cite{Grunthal_2024} and marked by the red dashed line.
These plots indicate that the ongoing timing campaign of  PSR J1618--3921
should be able to rule out the presence of 
Kozai resonance in this eMSP binary.}

\end{figure*}

\par 
It should be obvious that we now need an initial $e_2$ estimate while dealing with 
Eqs.~\ref{etgt_Eq} and~\ref{c2g12_Eq}, and the above inequalities. 
This is obtained by equating the general relativistic periastron precession timescale of the inner binary to the characteristic timescale for the Kozai-Lidov oscillations. 
We obtain  the period for general relativistic periastron precession using  
the relation $2\,\pi/ \dot \omega $ where $\dot \omega$ is given by $0.00142^\circ /yr$ as evident from 
Table 3 of \cite{Grunthal_2024}. 
Further, the period of Kozai oscillations is approximately described by
\begin{equation}
\label{t_Kozai}
 \tau_{\rm Kozai}  = P_i \, \frac{ M_i}{m_2}\, \left ( \frac{a_2}{a_1} \right )^3\, ( 1 - e_2^2 )^{3/2} \,,
\end{equation}
where $P_i$ is the orbital period of the inner binary~\cite{Mazeh_Shaham_1979}. 
Further, we note that an estimate for \( a_1 \), the semi-major axis of the MSP-HeWD binary, is obtained using Kepler's third law, and it leads to \( a_1 \approx 0.17\,\mathrm{AU} \) where we used the measured 
values \( \sim 22.7 \) days and \( \sim 1.42\,M_\odot \)
for $P_i$ and $M_i$, respectively.
This is indeed consistent with 
the observed projected semi-major axis \( x = 10.27\,\mathrm{lt\text{-}s} \) listed in Table 3 of~\cite{Grunthal_2024} if the orbital inclination angle \( i \approx 7^\circ \). 
This low inclination angle is consistent with the inferences available in \cite{Grunthal_2024}.

\par 
With the help of the above equations and expressions, we now model the proposed HT triple configuration in PSR J1618--3921.
It should be obvious that we need to provide estimates for $\alpha = (a_1/a_2)$, $e_2, $ and $\theta$ to pursue such a modeling. Clearly, we are allowed to let $m_0, m_1$ and $m_2$ be $1.2M_\odot, 0.2 M_\odot$ and $0.6M_\odot$ that provide the masses of the pulsar, its HeWD companion, and the third body, respectively, and we let the measured $e=0.027$ and $\omega=353^\circ$ to be the initial values of $e_1$ and $g_1$, respectively. To obtain admissible values for $\alpha$, we employ the relation \[
\frac{a_1}{a_2} = \left( \frac{P_1^2 \, (M_i+m_2)}{P_2^2 \, M_i} \right)^{1/3},
\] where we take $M_i\sim 1.42M_\odot$ and $m_2\sim 0.6M_\odot$,  which implies that $\alpha$ takes values 1/200, 1/320, 1/420, and 1/510 when we let the third body's orbital period be 150, 300, 450, and 600, respectively.
These considerations imply that we get $ e_2\sim 0.93$  when we let $\alpha\approx 1/320$ corresponding to $P_2\sim 300$ yr with the help of Eq.~\ref{t_Kozai}.
Finally, we let $\iota$ be in certain intervals depending on the scenarios we are trying to model.

\begin{figure*}[!hbt]
    \centering
    \includegraphics[width=0.95\textwidth]{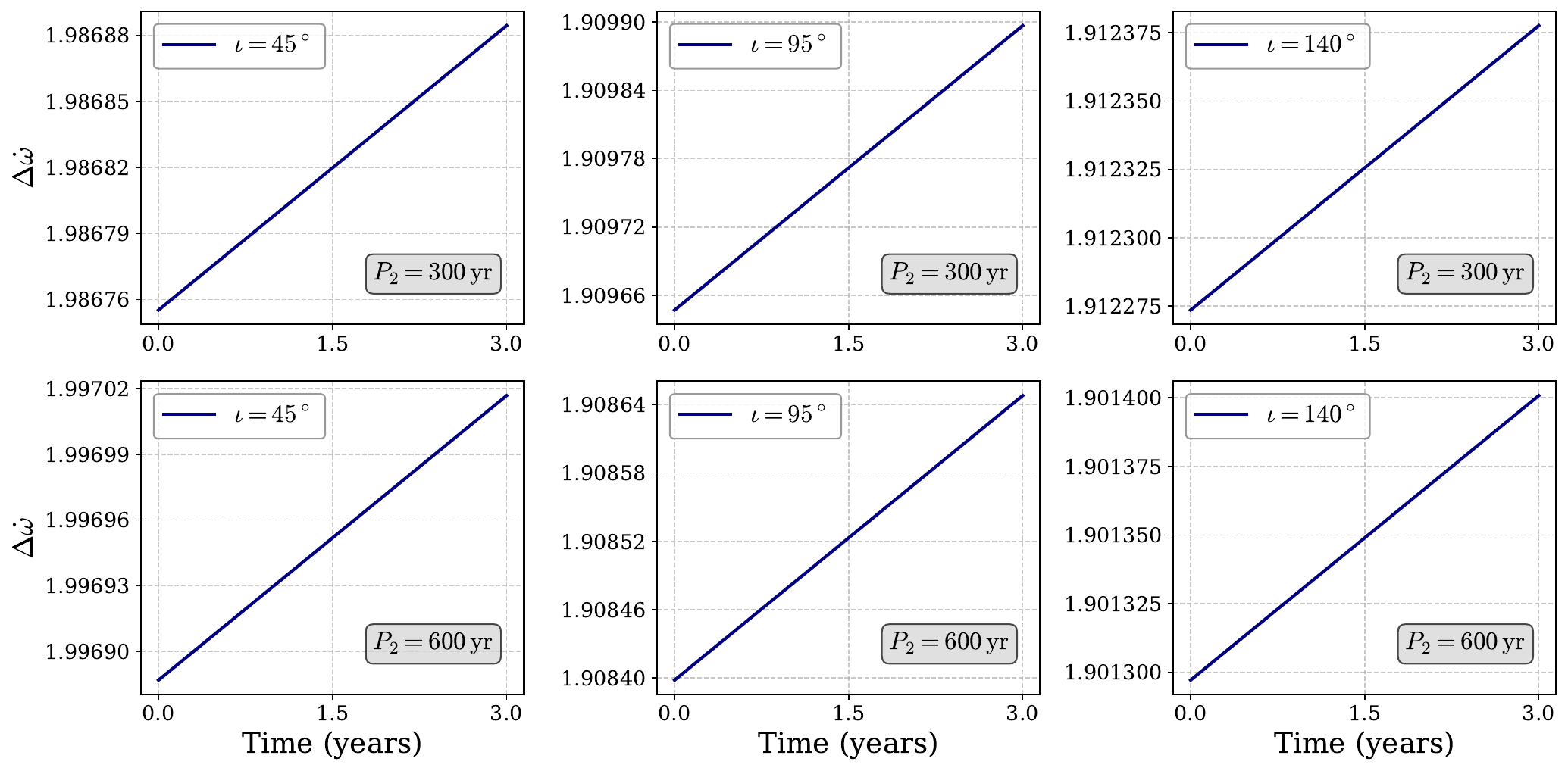}
    \caption{\label{fig:frac_change_gdot} 
     We plot fractional changes in the rate of periastron advance of our eMSP binary,
     namely  $\Delta \dot{\omega}(t) = \frac{\dot{g}_1(t) - \dot{\omega}}{\dot{\omega}}$ where 
     $ \dot{\omega} \sim 0.00142^\circ\, \mathrm{yr}^{-1}$, 
     as a function of time for scenarios depicted in Fig.~\ref{fig:frac_change_e1}. 
    The above  plots and those in Fig.~\ref{fig:frac_change_e1} indicate that a few additional months of 
MeerKat timing should allow one to rule out the presence of Kozai resonance in  PSR J1618--3921.
    }
\end{figure*}

\section{\label{sec3:level1}Observational implications of the Proposed  Hierarchical Triple Scenario for PSR~J1618--3921}

We now model a HT scenario for  PSR J1618--3921 while using observational and inferred parameters/constraints as listed in Table 3 of \cite{Grunthal_2024}.
We first explore the observational implications of a scenario where the proposed HT might be experiencing the Kozai resonance at the present epoch.
This consideration is influenced by the measured non-negligible (and unusually large) orbital eccentricity $e \sim 0.0274 $ of PSR J1618--3921.
For modeling a HT scenario for our eMSP binary, as noted earlier, we identify the measured orbital eccentricity, argument of periastron, and extracted semi-major axis values to be the initial values of variables like 
$e_1, g_1$ and $a_1$ that appear in Eqs.~\ref{etgt_Eq} and \ref{c2g12_Eq}. 
This implies that the only unknown variable is $\iota$, 
as the values of $e_2$, $a_2$, and $m_2$ are constrained 
by various considerations, detailed in the previous section.
To ensure that the proposed HT in  PSR J1618--3921 is experiencing Kozai resonance at the present epoch, we let $\iota$ vary between $39^\circ$ and $141^\circ$ influenced by \cite{Blaes_2002}.
 \par 
In Fig.~\ref{fig:frac_change_e1}, we plot fractional changes in the inner binary orbital eccentricity ($\Delta e_1 = (e_1(t) - e_1)/e_1$) as a function of time while identifying the initial value of $e_1$ to be 
$0.0274$, as given in \cite{Grunthal_2024}, and we
vary the orbital period of the 
 third body orbit between $150$yr and $600$ yr.
A close inspection of these plots reveals that the measured orbital eccentricity should decrease at the present epoch. More importantly, the changes in $e$ are substantially larger than the precision with which $e$ is measured by \cite{Grunthal_2024} as their measurement precision is around a few parts in $10^7$ while the expected temporal evolution is a few parts per $10^5$ as evident from Fig.~\ref{fig:frac_change_e1}.
{\it This implies that the ongoing MeerKat monitoring of PSR J1618--3921 under the ``RelBin'' project~\cite{Kramer_2021} should be able to rule out the possibility that this interesting eMSP binary is experiencing Kozai oscillations in the current epoch.
}

\par
We now explore how the measured rate of periastron advance varies if our eMSP binary is part of a HT that is experiencing Kozai oscillations. 
In Fig.~\ref{fig:frac_change_gdot}, we display plots for the fractional changes in the periastron precession rate, 
 namely $\Delta \dot{\omega}(t) = \frac{\dot{g}_1(t) - \dot{\omega}}{\dot{\omega}}$, where 
 $ \dot{\omega} \equiv 0.00142^{\circ}\text{yr}^{-1}$  for  $\iota$ and $\alpha$ values as in 
 Fig.~\ref{fig:frac_change_e1}. 
 An interesting aspect of these plots is that the rate of change of the inner binary's argument of periastron 
 is substantially different from the measured $\dot \omega$ value for our eMSP binary. 
 There could be two possibilities: i) the proposed HT configuration is not experiencing the Kozai resonance at the present epoch, and ii) the measured $\dot \omega$ could have classical (third-body) contributions 
 in addition to the general relativistic one.
 This is an important consideration as our employed $M_i$ value demands that the measured $\dot \omega$ arises only from General Relativity.
 Therefore, the ongoing timing campaign on PSR J1618--3921 should be able to clarify which of the two options suits the observations better.

\begin{figure*}[!ht]
    \centering
    \includegraphics[width=0.89\textwidth]{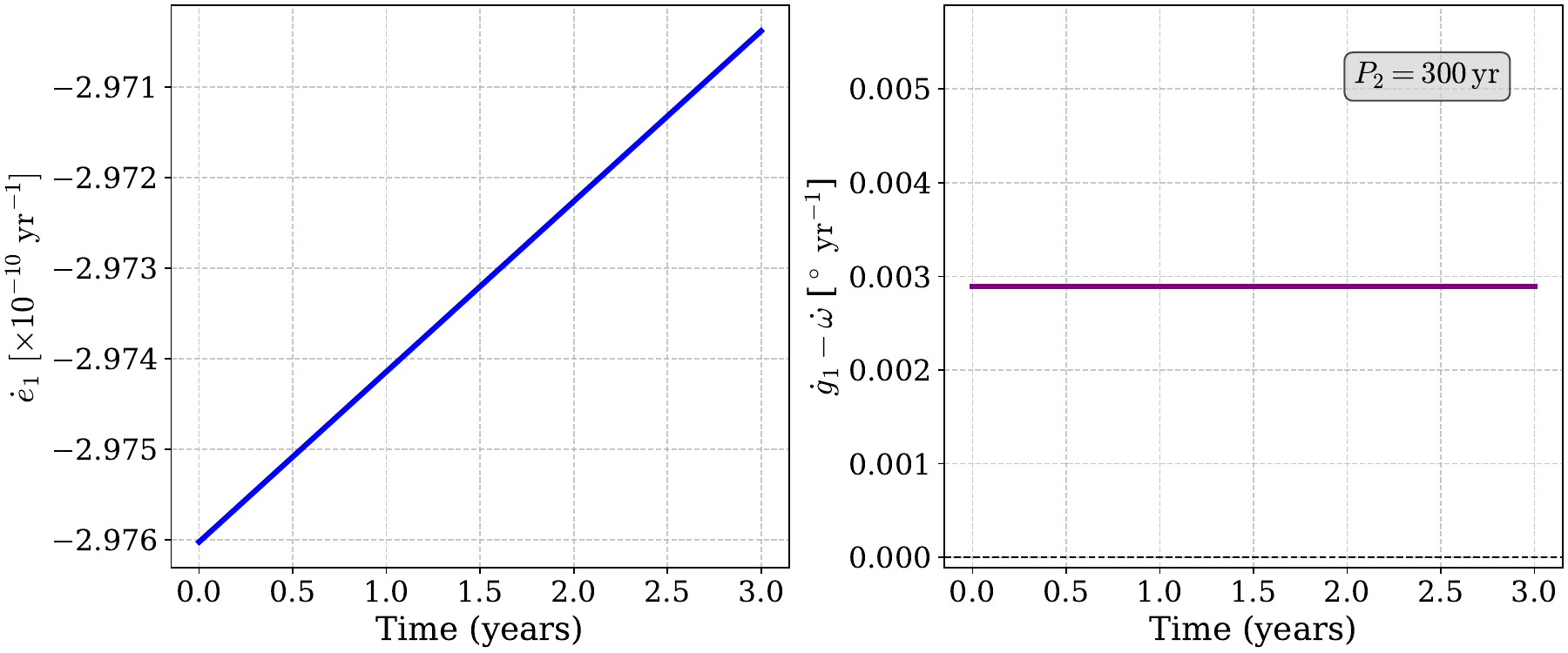}
    \caption{    
Plots that probe observational implications of  PSR J1618--3921 being part of a HT configuration. We choose $\iota$ so as to minimize the perturbations induced by the third body on 
the timed eMSP binary, which demands that $(\theta^2 = 1 - e_1^2) \rightarrow \iota \approx 1.1^\circ$. The predicted 
$\dot e_1$ for such an almost coplanar HT configuration is $\sim 10^{-10}$/yr and this is $\sim 10^5$ times larger than what is expected from General Relativity ($ \dot e_{\rm GR} \sim 10^{-15}$/yr).
The right panel plots are for $\dot g_1$ - $\dot \omega$ where 
$ \dot{\omega} \sim 0.00142^\circ\, \mathrm{yr}^{-1}$ as reported in \cite{Grunthal_2024}, and it clearly 
shows that even if PSR J1618--3921 is part of an almost coplanar HT configuration, the resulting  periastron advance rate should be different from the currently measured  
$\dot \omega$ value. These plots are for  $P_2\sim$ 300yr, and similar plots for $P_2\sim$ 600yr are 
rather indistinguishable from the ones displayed above, indicating that these estimates are largely insensitive to outer orbit parameters.
   }
    \label{fig:e1dot_g1dot}
\end{figure*}

\par
It may be pointed out that Ref.~\cite{Grunthal_2024} never argued that PSR J1618--3921 should be part of a HT that is currently experiencing Kozai oscillations.

This prompted us to probe scenarios where this eMSP binary is merely part of HT with specifications listed in \cite{Grunthal_2024} and those derived in the previous section, while ensuring that $\iota$ is not in 
the [$39^\circ-141^\circ$] range.
A closer examination of Eqs.~\ref{etgt_Eq} and~\ref{c2g12_Eq} shows that configurations minimizing the classical contributions to $\dot{g}_1$ and $\dot{e}_1$ arise under specific geometric conditions. In particular, minimizing the eccentricity evolution requires $\theta^2 = 1 - e_1^2$ as $\dot e_1 \propto (1 - \theta^2)$.

This implies that $\iota \sim 1^\circ$, corresponding to an almost coplanar HT configuration in which the dynamical influence of the outer binary is expected to be minimal, as evident from 
Eqs.~\ref{etgt_Eq} and \ref{c2g12_Eq}.
However, this configuration fails to minimize $\dot{g}_1$, as the second term in Eq.~\ref{dgdt}—which is highly sensitive to $\theta$—still makes a significant contribution.
In contrast, when $\theta \approx 0$, the term disappears altogether, yielding a separate configuration with $\iota \sim 90^\circ$, in which $\dot{g}_1$ reaches its minimum.

To investigate these complications, we first explore the temporal evolution of  $\dot e_1$ and $\dot g_1$ for the almost coplanar configuration
while allowing the third orbit to have an orbital period of either $\sim300$ or $\sim600$ years.

The left panel plots show that $\dot e_1$ variations are $\sim 10^{-10}$/yr and this is substantially larger than 
$10^{-15}$/yr due to the GW emission induced variation to $\dot e$ of
our MSP-HeWD binary \cite{BS89}. 
In other words, an actual measurement of the rate of change of timing eccentricity in 
PSR J1618--3921 should allow us to confirm if this eMSP binary is part of a HT.
However, this may be demanding as the measurement precision for $e$ is around a few
parts in $10^7$, which suggests a $\sim 10^3$ yr timing to constrain $e$ evolution to the
above $\dot e_1$ estimate. 
The forthcoming Square Kilometre Array (SKA), with its greatly enhanced sensitivity, is expected to deliver pulsar timing precision at the $\sim 10$ ns level for the best millisecond pulsars, thereby reducing the observational timescale required to reach such accuracy. 
It turns out that the combined precise measurements of $\dot{e}$ and $\dot{\omega}$ will be crucial for constraining the presence and properties of a potential third body in this intriguing eccentric MSP binary, as we discuss next.

\par

We infer from the $ \dot g_1 - \dot \omega $ plot in Fig.~\ref{fig:e1dot_g1dot} that the measured 
 $\dot \omega $ is more than a factor of two smaller than the estimated $ \dot g_1$ if
 our eMSP binary is part of a HT with $m_2 \sim 0.6\,M_\odot$ as required by ~\cite{Grunthal_2024}. 
 Recall that the plots in Fig.~3 are for configurations that minimize the evolution of  $\dot{e}_1$ due to the proposed third body in PSR J1618$-$3921, as we let $ \iota \sim 1^\circ$.
This discrepancy motivated us to explore whether varying $m_2$ in such nearly coplanar configurations could reconcile the difference. However, even after systematically adjusting $m_2$ and re-evaluating the relevant equations, the resulting values of $\dot{g}_1$ remained essentially unchanged.
A closer inspection of Eqs.~\ref{etgt_Eq} and \ref{c2g12_Eq} —particularly the coefficients ${\cal C}_2$, ${\cal G}_1$, and ${\cal G}_2$—reveals that $\dot{g}_1$ depends only weakly on $m_2$, owing to effective cancellations among these terms.
In other words, the configurations corresponding to minimum $\dot{e}_1$ values do not yield $\dot{g}_1$ estimates that are comparable to the measured $\dot{\omega}$ value.
\par 

This motivated us to explore creating configurations where $\dot{g}_1 = \dot{\omega}$. 
One may recall that the HT configurations for this eMSP binary essentially require only three parameters to specify the outer binary, and these are 
 i) mass of the third body $m_2$, ii) the mutual orbital inclination $\iota$, and 
 iii) the $ \alpha= a_1/a_2$ parameter. 
 To achieve the $\dot{g}_1 = \dot{\omega}$, we require an inclination angle $\iota$ close to $90^\circ$, while treating $m_2$ as a free parameter to probe the physically allowed space fully.

Under these conditions, equality between $\dot{g}_1$ and $\dot{\omega}$ can be realized for $\iota \approx 100^\circ$ and $m_2 \sim 0.01M\odot$.
However, a third body of such small mass is highly unlikely to produce the line-of-sight acceleration required to explain the observed $\dot{P}_b$.

These numerical experiments further reveal that the inclination angles minimizing the evolution of $\dot{g}_1$ and $\dot{e}_1$ are mutually exclusive, making it impossible to satisfy both constraints simultaneously. In other words, our results suggest that the proposed HT configuration for this intriguing eMSP system is not fully compatible with the current observational constraints reported by \cite{Grunthal_2024}.

\par 

This prompted us to pursue further numerical experiments 
such that $\dot g_1 $ expression, given by  Eq.~\ref{dgdt},
provides us the $ \dot \omega$ value as listed in Table 3 of \cite{Grunthal_2024}.
 It turns out that if we let $(a_2/a_1) > 4400$, the $\dot g_1 $ expression  indeed 
 provides the measured  $ \dot \omega$ value when we assign other parameters to values used 
 in our above displayed plots. 
 It is not very clear if such a configuration could be treated as a bound HT system in the 
 neighborhood of PSR J1618--3921.
 In a similar vein, we let $\theta$ be a free parameter so that the $\dot g_1 $ expression could lead to the measured $ \dot \omega$ value and our numerical experiments provide no real value solutions.
In another numerical experiment, we imposed the possibility that the measured $\dot \omega$ need not arise from the first post-Newtonian contribution to periastron advance as required by Eq.~6 in \cite{Grunthal_2024}.
This allowed us to treat $M_i$ as a free parameter while keeping other parameters at their appropriate values.
Unfortunately, we were unable to find real values of $M_i$ such that the above $\dot g_1$ expression 
gave us the measured  $ \dot \omega$ value.
These numerical experiments suggest that it is rather difficult to find an HT configuration 
for PSR J1618--3921 consistent with all the observations and inferences detailed in \cite{Grunthal_2024}.

\section{Conclusions}
\label{sec4}

We probed the observational consequence of the possibility that an interesting eMSP binary PSR J1618--3921 might be part of a HT configuration.
 Influenced by \cite{Grunthal_2024}, we let the inner binary consist of the timed MSP and its HeWD companion while the outer orbit involves a star of mass $0.6M_\odot$ with an orbital period $ \geq 300$ yrs.
 Adapting  \cite{Blaes_2002}, we model the system to be a point mass HT configuration under the influence of classical quadrupolar contributions and the leading-order general relativistic corrections to the rate of periastron advance.
 With these inputs, we show that orbital 
 $e$ and $\dot \omega$ should experience detectable temporal evolution in the coming years, especially if 
 the system is undergoing a Kozai resonance at the present epoch.
Additionally, 
we have shown that the HT configurations that minimize the temporal evolution of $e$ and $\dot \omega$
 are mutually incompatible. In practice, one can find HT setups in which 
 $\dot e$ 
 is negligible, but these predict a measurably different $\dot \omega$
 (and vice versa).
 Taken together, these considerations imply that continued high-precision timing of PSR J1618–3921—analyzed within the framework introduced here—should place stringent limits on the presence and properties of a potential third body in this intriguing eMSP binary.

\par
Very recently, we were informed that a change of even a few $\times 10^{-5}$ in orbital eccentricity would have been readily detectable in the existing dataset, and that a significant temporal evolution in $\dot \omega$ is also highly unlikely (Dr. Vivek Venkatraman Krishnan, private communication).

\begin{acknowledgments}
We thank the anonymous referee for valuable inputs, Vivek Venkatraman Krishnan, and Michael Kramer for helpful discussions.
  AG acknowledges the support of the Department of Atomic Energy, Government of India, under project identification \#RTI 4002.  PA
 acknowledges the support from SERB-DST, Govt. of India, via project
 code CRG/2022/009359.  
\end{acknowledgments}

\nocite{*}

\bibliography{PRD_ref}

\end{document}